\documentclass[twocolumn,prb,preprintnumbers,superscriptaddress,letterpaper,amsmath,floatfix,amssymb,showpacs]{revtex4}

\usepackage{graphicx}
\usepackage{times}
\usepackage{dcolumn}
\usepackage{natbib}
\usepackage{float}
\usepackage{comment}

\begin{document}

\preprint{}

\title{Magneto-Transport Properties of Single Crystalline LaFeAsO}

\author{C. A. McElroy}
\affiliation{Department of Physics, University of California, San Diego, La Jolla, California 92093, USA}
\affiliation{Center for Advanced Nanoscience, University of California, San Diego, La Jolla, California 92093, USA}

\author{J. J. Hamlin}
\altaffiliation{Current Address: Department of Physics, University of Florida, Gainesville, Florida 32611, USA}
\affiliation{Department of Physics, University of California, San Diego, La Jolla, California 92093, USA}
\affiliation{Center for Advanced Nanoscience, University of California, San Diego, La Jolla, California 92093, USA}

\author{B. D. White}
\affiliation{Department of Physics, University of California, San Diego, La Jolla, California 92093, USA}
\affiliation{Center for Advanced Nanoscience, University of California, San Diego, La Jolla, California 92093, USA}

\author{M. A. McGuire}
\affiliation{Materials Sciences and Technology Division, Oak Ridge National Laboratory}

\author{B. C. Sales}
\affiliation{Materials Sciences and Technology Division, Oak Ridge National Laboratory}

\author{M. B. Maple}
\altaffiliation{Corresponding Author: mbmaple@ucsd.edu}
\affiliation{Department of Physics, University of California, San Diego, La Jolla, California 92093, USA}
\affiliation{Center for Advanced Nanoscience, University of California, San Diego, La Jolla, California 92093, USA}

\date{\today}

\begin{abstract}

Measurements of magnetization, specific heat, electrical resistivity, Hall effect, and magnetoresistance on single crystalline samples of LaFeAsO grown in a NaAs flux are reported. While this material is known to be a semimetal, the temperature dependence of the electrical resistivity data presented herein is reminiscent of semiconducting behavior and exhibits distinct features associated with a structural transition and spin density wave (SDW) order. Magnetoresistance and Hall coefficient measurements were performed in magnetic fields up to 9 T applied perpendicular to the basal plane using a van der Pauw configuration. The charge carrier density and mobility indicate that electrons are the majority charge carriers and exhibit features indicative of the structural transition and SDW formation. Low temperature X-ray diffraction measurements have confirmed that the structural transition in these samples occurs near 140 K, compared to a transition temperature of 156 K observed in polycrystalline samples. Isotherms of magnetoresistivity measured as a function of magnetic field can be scaled onto a single curve in which the scaling field is a linear function of temperature between 2.2 K and 180 K.

\end{abstract}


\maketitle

\section{INTRODUCTION}

The superconductivity observed in the fluorine substituted layered pnictides LaFePO and LaFeAsO,\cite {KAMIHARA01,KAMIHARA02} generated an enormous amount of interest in these and other iron pnictide and chalcogenide compounds.\cite{ASWATHY01} This interest is primarily driven by the correlated electron phenomena these compounds display such as high-$T_{c}$ superconductivity, commensurate and incommensurate spin-density wave order, and the interplay of these phenomena, as well as structural transitions and pressure-induced isostructural volume collapses.\cite{SI01,JOHANNES01,FERNANDES01,PRATT01,MCGUIRE01,FU01,LIU01} The highest $T_{c}$'s among the iron pnictide superconductors have been found in the arsenide 1111 compounds, such as Gd$_{1-x}$Th$_{x}$FeAsO with an onset of $T_{c}$ = $56$ K when $x = 0.20$.\cite{WANG01} There are also structural and electronic similarities with the cuprates where the FeAs layers govern the electronic states near the Fermi level in a manner analogous to the role played by the CuO$_{2}$ layers in cuprate materials.\cite{KIVELSON01}

Most studies to date on the FeAs-1111 system have been conducted on polycrystalline samples which were synthesized via solid state reaction. Although single crystals of phosphorus based 1111s have been studied in some detail, \cite{HAMLIN01,HILLIER01,FLETCHER01} relatively few single crystal studies of \textit{Ln}FeAsO (\textit{Ln} = lanthanide) compounds have been carried out, leading to a clear gap in knowledge concerning the landscape of FeAs-1111 physics. The development of suitable fluxes including NaAs, \cite{YAN01} and KI, \cite{JESCHE01} for the growth of \textit{Ln}FeAsO compounds has provided the impetus for studies of single crystal samples of this class of materials. Flux growth techniques provide a reasonably accessible method for growing \textit{Ln}FeAsO single crystals, enabling studies of crystals that are large enough to examine intrinsic properties in greater detail as well as to characterize and study the anisotropy of various physical properties. The ability to control crystalline orientation in the case of this tetragonal/orthorhombic material makes it possible to characterize the anisotropy of the electrical and magnetic properties\cite{JESCHE01,YAN01} and provides insight into the more subtle electronic behavior of the material.

In this paper, we report magnetotransport measurements on single crystalline samples of LaFeAsO. From an analysis of Hall coefficient and magnetoresistivity measurements, it is possible to extract the charge carrier concentration and mobility. Pallecchi \textit{et al.} have recently performed magnetoresistivity measurements on polycrystalline samples of LaFe$_{1-x}$Ru$_{x}$AsO and find a linear term in the magnetoresistance curves which they attribute to the presence of Dirac cones in the band structure.\cite{PALLECCHI01} Magnetoresistance measurements on the single crystalline specimens of LaFeAsO reported herein over a wide temperature range and in magnetic fields up to 9 T do not exhibit a similar linear contribution; instead, they reveal a simple and systematic behavior which is amenable to scaling the magnetoresistivity isotherms onto a universal curve.

\section{EXPERIMENTAL DETAILS}

Single crystals of LaFeAsO were synthesized in a Ta tube using a molten NaAs flux.\cite{YAN01}  NaAs was synthesized by mixing stoichiometric quantities of Na metal and As lumps (Alfa Aesar, 99.999\%) in a Ta tube (0.500" O.D. $\times$ 8") in a glove box under an Ar atmosphere. The Ta tube was arc-welded shut under approximately one atmosphere of Ar in an arc furnace and double sealed in evacuated quartz ampoules. The NaAs charge was heated to 565 $^{\circ}$C for $15$ hrs., cooled to room temperature, and removed again in an Ar atmosphere. LaAs was synthesized by mixing stoichiometric quantities of La powder (Alfa Aesar, 99.9\%, -40 mesh) and As lumps and double sealing them in evacuated quartz ampoules. The charge was placed horizontally in a furnace to provide a greater exposed surface area in order to promote a thorough reaction. The furnace was heated to 300 $^{\circ}$C over $10$ hrs., dwelled at 300 $^{\circ}$C for 10 hrs., heated to 600 $^{\circ}$C over 10 hrs., dwelled at 600 $^{\circ}$C for 10 hrs., and heated to 800 ${^\circ}$C over 4 hrs., dwelled 800 ${^\circ}$C for 12 hrs., before cooling to room temperature and removing the charge from the quartz ampoules under an Ar atmosphere. The long dwell at 600 $^{\circ}$C during the warming curve addresses safety concerns regarding the sublimation of As and the exothermic reaction with La. LaFeAsO was synthesized by mixing the LaAs precursor, Fe$_{2}$O$_{3}$ (99.998\%), Fe powder (99.998\%) and the NaAs flux in a 1:20 ratio (charge:flux) in a tantalum tube ($0.500$" O.D. $\times$ $8$") while working in an Ar glove box. The tube was arc-welded shut, double sealed inside quartz ampoules and heated in a top loading furnace to 1150 ${^\circ}$C over 37.5 hrs., dwelled at 1150 ${^\circ}$C for 24 hrs., and cooled to 600 ${^\circ}$C over 183 hrs. The furnace was shut off at 600 $^{\circ}$C to allow the contents of the Ta tube to cool rapidly to room temperature. The charge and flux were carefully removed from the tantalum tube in an Ar atmosphere.

\begin{figure}[t]
\begin{center}
    \includegraphics[scale=0.315]{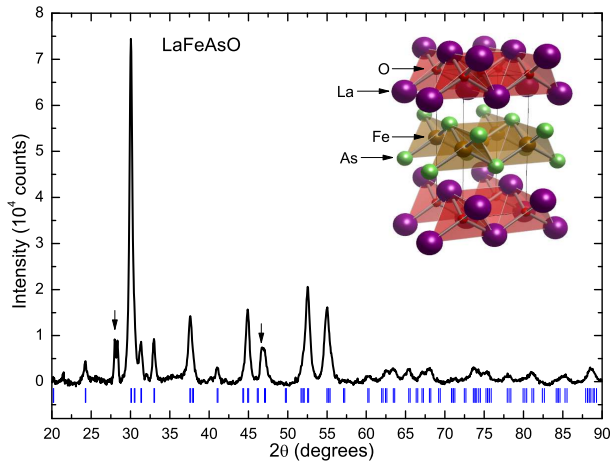}
    \caption{Powder X-ray diffraction data for flux-grown single crystals of LaFeAsO. A La$_{3}$TaO$_{7}$ impurity phase was observed (indicated by peaks identified with arrows) which comprises approximately 2\% of the molar fraction of the sample. The inset dispalys the ZrCuSiAs-type crystal structure with tetragonal space group $P4/nmm$.}
    \label{fig:CAMLaFeAsOXRD01}
\end{center}
\end{figure}

\begin{figure}[t]
\begin{center}
    \includegraphics[scale=0.315]{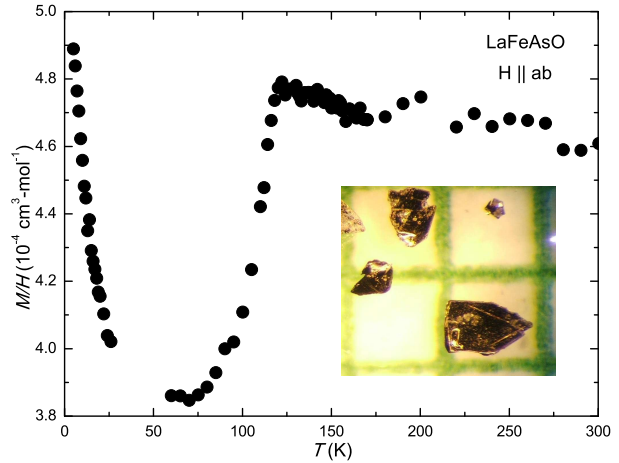}
    \caption{Magnetization $M$ divided by magnetic field $H$ vs. temperature, measured after zero-field cooling with the $ab$ plane oriented parallel to a 5.5 T applied field. The transition to spin density wave order near 118 K is clearly evident. The inset shows a photograph of several typical single crystals superimposed on graph paper with a 1 mm $\times$ 1 mm grid.}
    \label{fig:CAMLaFeAsOchi01}
\end{center}
\end{figure}

The growth yielded a large quantity of crystalline platelets, with dimensions of up to 1.5 $\times$ 1.5 $\times$ 0.05 mm$^{3}$. A photograph of several typical crystals is shown in the inset of Fig.~\ref{fig:CAMLaFeAsOchi01}. Residual NaAs flux was removed by a delicate etch in deionized water for approximately 30 s. The LaFeAsO crystals are very resilient to oxidation in air and robustly tolerate applied stress despite their thinness; it is possible, however, to damage the yield with a vigorous etch if too much material is washed at once. Crystals were sorted from the removed material and carefully introduced into the deionized water to avoid this problem.

Powder X-ray diffraction (XRD) measurements were performed on etched and powdered crystals. Using a Bruker D8 Discover diffractometer, the powder was scanned at room temperature from 20${^\circ}$ to 90${^\circ}$ in $2\theta$ with a Cu$_{K\alpha}$ rotating anode target. Four crystals were examined with Energy Dispersive X-ray spectroscopy (EDX) measurements at three separate locations on each crystal to confirm the stoichiometry of elements on the surface of the sample. Low temperature XRD measurements were performed at Oak Ridge National Laboratory (ORNL) with a PANalytical X'pert Pro Multi-Purpose Diffractometer (MPD) using monochromatic Cu K$_{\alpha1}$ radiation and an Oxford PheniX closed cycle cryostat. For the low temperature measurements, special attention was payed to the (2,2,0) tetragonal peak as it splits into the (4,0,0) and (0,4,0) orthorhombic peaks through the structural transition, as described in Section III, Results and Discussion.

The X-ray diffraction pattern, measured from powdered LaFeAsO crystals at room temperature, is shown in Fig.~\ref{fig:CAMLaFeAsOXRD01}. The pattern confirms the previously reported ZrCuSiAs-type crystal structure with $P4/nmm$ space group. Small (La$_{3}$TaO$_{7}$) impurity peaks were observed and are attributed to reactions of the starting materials with the Ta tube during firing. Rietveld refinement of the x-ray data was performed using GSAS,\cite{LARSON01} which yielded  lattice constants $a = b = 4.0367(4)$ {\AA} and $c = 8.793(1)$ {\AA}. Reported values for the lattice parameters of LaFeAsO \cite{QUEBE01} differ by approximately 0.02\% in the $ab$ plane and approximately 0.4\% for $c$. The stoichiometric ratio of the single crystals was confirmed to be 1:1:1:1 within experimental uncertainty by EDX measurements.

\begin{figure}[t]
\begin{center}
    \includegraphics[scale=0.315]{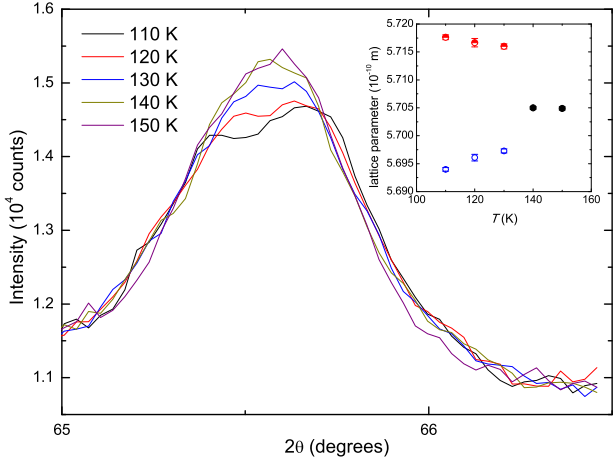}
    \caption{Powder X-ray diffraction data in the vicinity of the tetragonal (2,2,0) peak of LaFeAsO measured at temperatures between 110 K and 150 K show the splitting of the $a = b$ lattice parameters as the crystal structure transforms from the tetragonal to the orthorhombic phase. The inset shows the evolution of these lattice parameters with temperature, where the half-filled red circles and the open blue circles represent the $a$ and $b$ orthorhombic lattice constants, respectively, and the solid black circles represent the $a$$\sqrt2$ tetragonal lattice constant.}
    \label{fig:CAMLaFeAsOXRD02}
\end{center}
\end{figure}

\begin{figure}[t]
\begin{center}
    \includegraphics[scale=0.315]{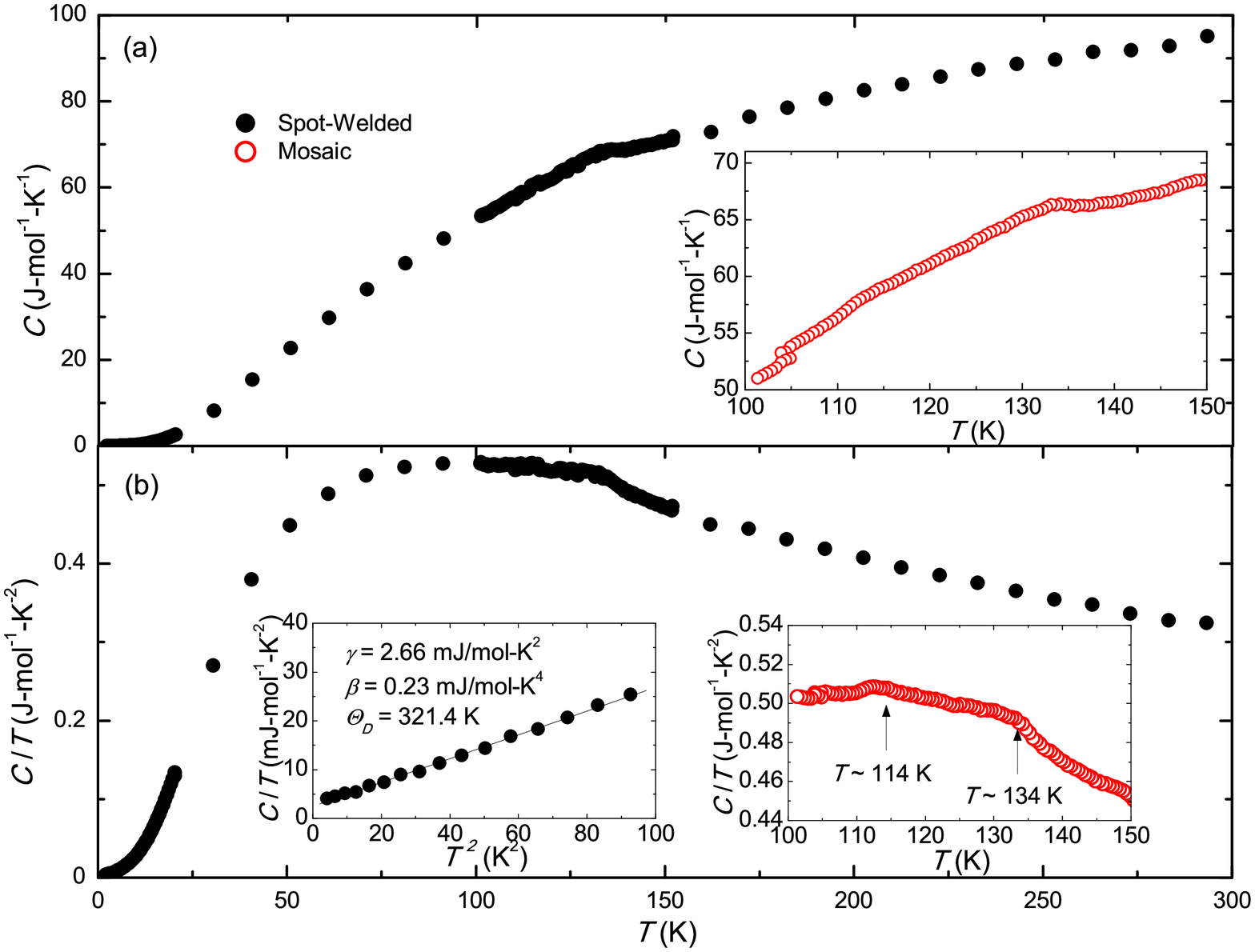}
    \caption{(a) Specific heat $C$ vs. temperature $T$ of single crystals of LaFeAsO where black filled circles are from measurements of a mosaic of spot-welded crystals and red open circles are from measurements of a mosaic comprised of a single layer  of crystals. The inset shows $C$ vs. $T$ data near the SDW and structural transition. (b) $C/T$ vs. temperature $T$. Displayed in the left inset is a low temperature fit of the expression $C/T$ = $\gamma$ + $\beta$$T^{2}$ to the data on a $C/T$ vs. $T^{2}$ plot. Values for $\gamma$ of 2.66 mJ/mol-K$^{2}$ and $\beta$ of 0.23 mJ/mol-K$^{4}$ were obtained, where the value of $\beta$ corresponds to a Debye Temperature $\Theta_{D}$ of 320 K. The right inset shows two features near the spin density wave and structural transition temperature in the $C/T$ vs. $T$ plot at 114 K and 134 K, respectively.}
    \label{fig:CAMLaFeAsOCp01}
\end{center}
\end{figure}

A mosaic of 10 crystals with a total mass of 1.77 mg was assembled for magnetization measurements using a Quantum Design SQUID Magnetic Properties Measurement System (MPMS). A magnetic field of 5.5 T was applied parallel to the $ab$ plane of the crystals for a zero-field cooled temperature sweep from 5 K to 300 K.

Specific heat measurements were made using a Quantum Design PPMS Dynacool with a standard thermal relaxation technique. Numerous single crystals with a combined mass of 1.24 mg were spot-welded together. This allowed a larger cumulative sample mass to be measured than would be possible by covering the 3 $\times$ 3 mm$^{2}$ specific heat platform with a single layer of crystals, without sacrificing good thermal contact to the platform. To verify that the specific heat measurement was not affected by spot-welding, a standard mosaic was also measured (with all samples in direct contact with the sample platform). The two methods yielded comparable results.

The electrical transport properties of LaFeAsO were measured using the van der Pauw method.\cite{VANDERPAUW01} The geometric flexibility of the van der Pauw method enables measurements of electrical transport properties to be made on thin samples with irregular shapes.  After sputtering gold pads on four distinct edges of a platelet-like single crystal, platinum wires were affixed to the sample using a two-part silver epoxy. Magnetoresistivity and Hall resistivity were measured on the same sample.

\begin{figure}[t]
\begin{center}
    \includegraphics[scale=0.315]{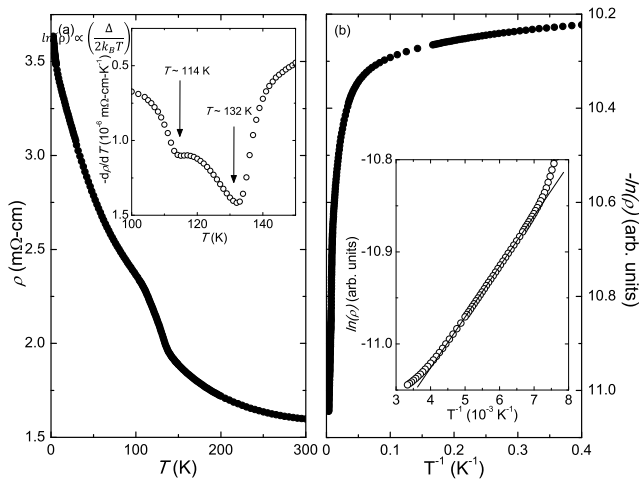}
    \caption{(a) Electrical resistivity $\rho$ vs. temperature $T$ of single crystal LaFeAsO. The derivative of the electrical resistivity with respect to temperature $d\rho/dT$, vs. $T$, shown in the inset, exhibits two clear features near 114 K and 132 K reflecting the spin density wave and structural transitions, respectively. (b) A plot of $\ln$$\rho$ vs. $1/T$ shows activated behavior above 132 K.  Displayed in the inset is a fit of an Arrhenius law $\ln\rho \propto \left ( \Delta /2k_{B}T \right )$ to the high temperature region above the SDW feature from which an energy gap of  $\Delta \simeq 9.5 \pm 0.03$ meV has been inferred.}
    \label{fig:CAMLaFeAsOrho01}
\end{center}
\end{figure}

The four platinum wires were assigned $I_{i}^{\pm}$ and $V_{i}^{\pm}$ for six distinct contact arrangements, each yielding effective resistances $R_{i}^{\pm}=V_{i}^{\pm}/I_{i}^{\pm}$ where $(i=1,2,...,6)$. Four of these arrangements, constituting two pairs, have their electrodes in adjacent positions. For these arrangements, averaging $\rho^{\pm}_{HA}$ and $\rho^{\pm}_{HB}$ from Eqs.~(\ref{equ:CAMLaFeAsOrhoEQ01}) and (\ref{equ:CAMLaFeAsOrhoEQ02}) in a positive and negative magnetic field enables the even-functioned magnetoresistance to be extracted from the observed signal. The two remaining arrangements have their electrodes positioned across from each other. Equations~(\ref{equ:CAMLaFeAsORHEQ01}) and (\ref{equ:CAMLaFeAsORHEQ02}) are combined to eliminate the even-functioned magnetoresistance, leaving the odd-functioned Hall resistance $R_{H}$. The quantities $f_{A}$ and $f_{B}$ are calculated using a transcendental equation involving ratios of resistances in each of the four magnetoresistance arrangements.

\begin{equation}
    \rho^{\pm}_{HA} = \frac{\pi f^{\pm}_{A} t}{\ln\left(2\right)}\left[\frac{\left(V^{\pm}_{1} + V^{\pm}_{2}\right)}{\left(I^{\pm}_{1} + I^{\pm}_{2}\right)}\right]
    \label{equ:CAMLaFeAsOrhoEQ01}
\end{equation}

\begin{equation}
    \rho^{\pm}_{HB} = \frac{\pi f^{\pm}_{B} t}{\ln\left(2\right)}\left[\frac{\left(V^{\pm}_{3} + V^{\pm}_{4}\right)}{\left(I^{\pm}_{3} + I^{\pm}_{4}\right)}\right]
    \label{equ:CAMLaFeAsOrhoEQ02}
\end{equation}

\begin{equation}
    R_{HA} = \frac{t}{B}\left[\frac{\left(R_{5}^{-} - R_{5}^{+}\right)}{2}\right]
    \label{equ:CAMLaFeAsORHEQ01}
\end{equation}

\begin{equation}
    R_{HB} = \frac{t}{B}\left[\frac{\left(R_{6}^{-} - R_{6}^{+}\right)}{2}\right]
    \label{equ:CAMLaFeAsORHEQ02}
\end{equation}

\section{RESULTS AND DISCUSSION}

Magnetization measurements performed as a function of temperature on single crystal LaFeAsO are plotted in Fig.~\ref{fig:CAMLaFeAsOchi01} and clearly reveal SDW order below 118 K.\cite{KLAUSS01} A study by Yan \textit{et al.} shows an enhanced magnitude of $M/H$ when $H$ is applied parallel to the basal plane; thus, to compensate for the small mass of our samples, we measured the magnetization for $H \parallel ab$. At temperatures below the structural transition, we might expect to see Curie-Weiss behavior due to a reduction of the Pauli paramagnetism resulting from localization of the charge carriers. However, recent NMR studies of LaFeAsO below the structural transition at 156 K show a cessation of the growth of orthorhombic domains by 120 K and present evidence for the emergence of a spin nematic phase.\cite{FU01} The low temperature upturn in $M/H$ (see Fig.~\ref{fig:CAMLaFeAsOchi01}) has been reported numerous times and is typically attributed to paramagnetic impurities.\cite{KAMIHARA02,MCGUIRE01} Our attempts to characterize this contribution with a Curie-Weiss fit were unsuccessful; measurements at lower temperature would be necessary to perform such an analysis. At higher temperature above the antiferromagnetic transition in LaFeAsO, there is little temperature dependence of $M/H$ in agreement with previous reports from polycrystalline \cite{KAMIHARA02,MCGUIRE01} and single crystalline \cite{FU01,CHEN01} samples.

XRD measurements performed at ORNL provided solid evidence that the structural transition in our single crystals occurs at a notably lower temperature than has been reported in other studies. Measurements from $20^{\circ}$ to $75^{\circ}$ at 298 K and 50 K (not shown) confirmed the presence of a structural transition from the high temperature tetragonal $P4/nmm$ space group to the orthorhombic $Cmma$ space group. The (2,2,0) tetragonal peak near $66.5^{\circ}$ was examined closely in the temperature range from 110 K to 150 K to evaluate more precisely the structural transition temperature and the results are shown in Fig.~\ref{fig:CAMLaFeAsOXRD02}. These data were fit using GSAS for both tetragonal and orthorhombic symmetries. The lattice parameters that were calculated show a clear splitting of the $a$ and $b$ lattice parameters as can be seen in the inset of Fig.~\ref{fig:CAMLaFeAsOXRD02}. XRD measurements at low temperature revealed the structural transition in our single crystals to occur near 140 K. The structural transition and SDW order occur at substantially lower temperatures in these crystals than has been reported previously for polycrystalline samples.\cite{MCGUIRE01}

Specific heat measurements on our LaFeAsO single crystals reveal two distinct features near 114 K and 134 K, as shown in the $C/T$ vs. $T$ data displayed in Fig.~\ref{fig:CAMLaFeAsOCp01}. These specific heat data corroborate the transition temperatures observed in our magnetization and low temperature XRD measurements. Studies on polycrystalline samples of LaFeAsO have shown a feature near 140 K associated with the SDW transition and a feature near 156 K due to the structural transition from tetragonal to orthorhombic symmetry.\cite{MCGUIRE01} The specific heat at room temperature is quite close to the expected Dulong-Petit value per mol LaFeAsO of $12R =$ 99.8 J/mol-K.

The Sommerfeld coefficient, $\gamma$, was determined from a linear fit of the relation $C/T = \gamma + \beta T^2$ to the $C/T$ vs. $T^{2}$ data at low temperature. From the phonon contribution ($\beta$) to $C(T)$, the Debye temperature ($\Theta_{D}$) was calculated using the relation $\beta = 1944n/\Theta_{D}^{3}$. It is worth mentioning that a small $\gamma$ is typical of classical semimetals, especially bismuth with a $\gamma = 67.0 \mu$J/mol K,\cite{COLLAN01} and that the low density of electronic states near the Fermi level is consistent with the small value of $\gamma$ = 2.66 mJ/mol-K$^{2}$ calculated from our data. Our value for the Debye temperature, $\Theta_{D}$ = 320 K, agrees with values reported elsewhere and our value for $\gamma$ is consistent with values of $\gamma$ reported in studies of polycrystalline specimens of LaFeAsO.\cite{MCGUIRE01} Using a similar approach as reported by McGuire \textit{et al.},\cite{MCGUIRE01} we estimated the entropy associated with the features in the right hand inset of Fig.~\ref{fig:CAMLaFeAsOCp01}(b). By approximating the background with a polymonial fit to the region above and below the transition and integrating the remaining $C/T$ data, we obtained $\Delta S \approx$ 0.37 J/K which is quite close to the 0.27 J/K reported by McGuire \textit{et al.}.\cite{MCGUIRE01} Efforts were made to estimate the background with a simple Debye model for $C(T)$. However, the phonon dispersion for LaFeAsO has been measured,\cite{CHRISTIANSON01} and is too complex to be adequately approximated by a linear dispersion relation.

Electrical resistivity $\rho(T)$ data for single crystalline LaFeAsO, measured using the van der Pauw method,\cite{VANDERPAUW01} are displayed in Fig.~\ref{fig:CAMLaFeAsOrho01}(a). While this material is widely regarded to be a semimetal, the temperature dependence of the electrical resistivity reported herein suggests activated semiconducing-like behavior from the SDW feature at 130 K to room temperature. A clear feature in $\rho(T)$ is observed at 114 K as is shown in the inset of Fig.~ \ref{fig:CAMLaFeAsOrho01}(a). This feature is consistent with the feature observed in specific heat measurements and is associated with SDW order. In contrast, the electrical resistivity of polycrystalline LaFeAsO decreases as the temperature is decreased through the SDW transition.\cite{MCGUIRE01} The occurrence of two distinct temperature dependencies of electrical resistivity, one of which shows a decrease in resistivity below 138 K like the polycrystalline data, has been reported in studies of other single crystals of LaFeAsO.\cite{YAN01} However, we were never able to observe this behavior in measurements of 15 distinct single crystals; instead, the crystals we have measured all appear to exhibit semiconducing-like behavior. An energy gap was extracted by fitting an Arrhenius law ( $\ln\rho \propto \left ( \Delta /2k_{B}T \right )$) to the -$\ln\rho$ vs $1/T$ data shown in Fig.~\ref{fig:CAMLaFeAsOrho01}(b) at high temperature (from 137 K to 300 K). This analysis yields an energy gap with magnitude $\Delta \sim$ 9.5 meV, which is similar to the behavior of some of the single crystals studied by Yan \textit{et al.}.\cite{YAN01}

\begin{figure}[t]
\begin{center}
    \includegraphics[scale=0.315]{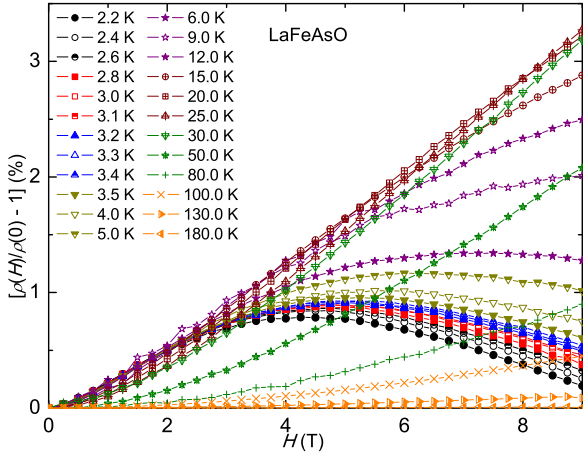}
    \caption{Isotherms of magnetoresistivity $\rho (H)/\rho (0)-1$ vs. applied magnetic field $H$ at temperatures between 6 K and 180 K. A broad maximum in the low temperature data shifts systematically to higher magnetic field with increasing temperature.}
    \label{fig:CAMLaFeAsOrhomag01}
\end{center}
\end{figure}

\begin{figure}[t]
\begin{center}
    \includegraphics[scale=0.315]{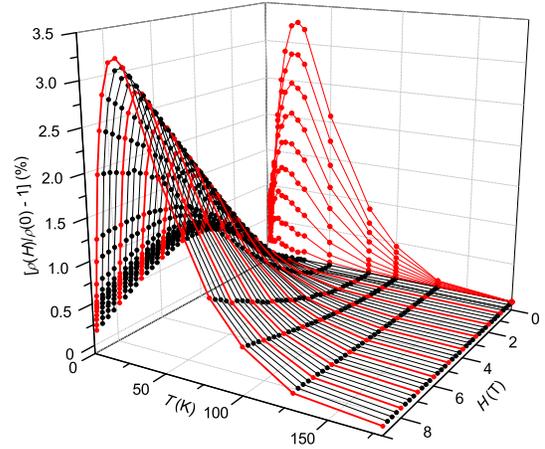}
    \caption{Isotherms of magnetoresistivity $\rho (H)/\rho (0)-1$ plotted as a function of both temperature $T$ and magnetic field $H$, showing the evolution of a maximum. Note that the isochamps (constant magnetic field) are displayed as projections from selected magnetic fields ($H = 1, 2, 3, ..., 9$ T) onto the magnetoresistivity-$T$ plane.}
    \label{fig:CAMLaFeAsOrhomag02}
\end{center}
\end{figure}

We measured the magnetoresistivity ($\rho(H)/\rho(0) - 1$) of LaFeAsO single crystals at various temperatures as shown in Fig.~ \ref{fig:CAMLaFeAsOrhomag01}. The magnetoresistivity curves show a continuous evolution in behavior from 2.2 K up to 180 K. At low temperatures, $\rho(H)/\rho(0) - 1$ reaches a maximum at modest fields and decreases above this maximum. The evolution of the maxima in $\rho (H)/\rho (0)-1$ as a function of both temperature and magnetic field is highlighted in Fig.~\ref{fig:CAMLaFeAsOrhomag02}. It is striking how much the maxima in the magnetoresistivity isotherms change whereas the maxima seen in the isochamps (constant magnetic field curves) are much more stable in fields up to 9 T.

\begin{figure}[t]
\begin{center}
    \includegraphics[scale=0.315]{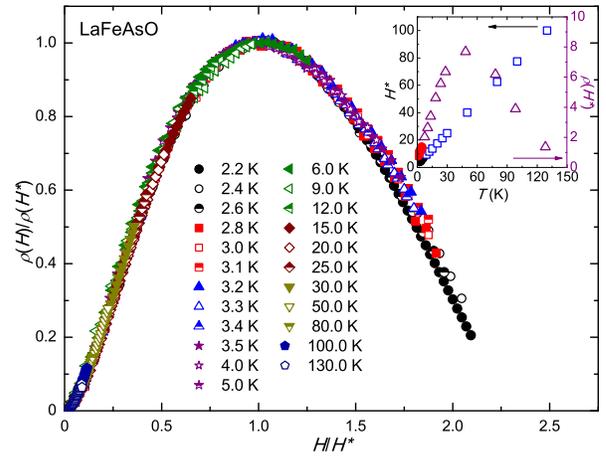}
    \caption{Scaled magnetoresistivity isotherms ($\rho (H)/\rho (0)-1$)/$\rho(H^{\ast})$ vs. reduced magnetic field $H$/$H^{\ast}$, where $\rho(H^{\ast})$ is the maximum value of the magnetoresistivity and $H^{\ast}$ is the value of $H$ at which the maximum occurs for isotherms between 2.2 K and 6 K.   For isotherms at temperatures greater than 6 K, values of $H^{\ast}$ were derived from a linear extrapolation of the values of $H^{\ast}$ between 2.2 K and 6 K (indicated by open squares in the inset of the figure), whereas values of $\rho(H^{\ast})$ were selected to bring the isotherms into coincidence with the isotherms between 2.2 K and 6 K (denoted by the open triangles in the inset to the figure).}
    \label{fig:CAMLaFeAsOrhomag03}
\end{center}
\end{figure}

Magnetoresistance measurements have recently been performed on polycrystalline LaFeAsO samples.\cite{PALLECCHI01}  These studies reveal a linear contribution to the magnetoresistance of LaFeAsO that is attributed to the presence of Dirac cones in the Fermi surface whose anisotropy is realized along the $\Gamma -$ X and $\Gamma -$ Y directions.  However, the magnetoresistance data reported herein on LaFeAsO single crystals do not show this linear contribution for similar temperatures and magnetic fields. The reason for this difference in behavior of polycrystalline and single crystalline samples is not clear. One possibility is that the magnetoresistance measurements on polycrystalline sample are an average of crystallites with varying orientations and anisotropic magnetoresistance.

Analysis of the temperature and magnetic field dependence of the magnetoresistance for LaFeAsO can be analyzed with the well known Kohler's rule,\cite{KOHLER01}

\begin{equation}
    \frac{R\left(H,T\right)}{R\left(0,T\right)}=f\left(\frac{H}{R\left(0,T\right)}\right).
    \label{equ:CAMLaFeAsOkohler}
\end{equation}

This kind of scaling is predicted to hold for materials with a single species of charge carrier and constant scattering rate at all points on the Fermi surface. Our single crystals of LaFeAsO do not appear to follow Kohler's rule (plot not shown), which is likely a consequence of the complex band structure of LaFeAsO and the presence of more than one species of charge carrier as will be addressed later in this paper.\cite{KRUGER01} Additional discussion of situations that can result in deviations from Kohler's rule are discussed in Ref.~\onlinecite{MCKENZIE01}.

Each isotherm in the magnetoresistivity data presented in Fig.~\ref{fig:CAMLaFeAsOrhomag01} can be fit well with a sixth order polynomial. This fit allowed us to identify the extrema $H^{\ast}$ and $\rho(H^{\ast})$ in $\rho (H)/ \rho (0)-1$, which are plotted in the inset of Fig.~\ref{fig:CAMLaFeAsOrhomag03}. Those magnetoresistivity isotherms that exhibit a maximum can be scaled to a common curve upon normalizing by $H^{\ast}$ and $\rho(H^{\ast})$ as shown in Fig.~\ref{fig:CAMLaFeAsOrhomag03}. That these data can be scaled in such a way indicates that the physics governing the electrical transport under magnetic field is qualitatively unchanged from 2.2 K to at least 130 K. Isotherms that do not display a maximum can be normalized in magnetic field with values extrapolated linearly from known $H^{\ast}$ values and scaled in magnitude until they lie on the curve. Values of $H^{\ast}$ and $\rho(H^{\ast})$ used to scale the magnetoresistivity data are plotted in the inset of Fig.~\ref{fig:CAMLaFeAsOrhomag03}.

In low magnetic fields, the cyclotron magnetoresistivity of a compensated material with two bands can be fit, in general, by Eq.~\ref{equ:CAMLaFeAsOrhomagEQ02} where $\alpha$ and $\beta$ are fitting parameters.\cite{PALLECCHI01} Thus, we expect our data to exhibit a quadratic magnetic field dependence predicted by Eq.~\ref{equ:CAMLaFeAsOrhomagEQ02}, but we observe that the range over which this behavior is obeyed depends on temperature. By plotting the scaled magnetoresistivity in Fig.~\ref{fig:CAMLaFeAsOrhomag03} verses $H^{2}$ (not shown), we were able to derive a relation for the maximum field ($H_{p}$) as a function of temperature in which the magnetoresistance exhibits a quadratic field dependence: $H_{p} = 0.76 + 0.22T$. Above $H_{p}$, the magnetoresistance cannot be described by a simple cyclotron contribution alone.

\begin{equation}
    \frac{\rho(H)-\rho(0)}{\rho(0)}=\frac{\alpha\left(\mu_{0}H\right)^{2}}{\beta+\left(\mu_{0}H\right)^{2}}
    \label{equ:CAMLaFeAsOrhomagEQ02}
\end{equation}

\begin{figure}[t]
\begin{center}
    \includegraphics[scale=0.315]{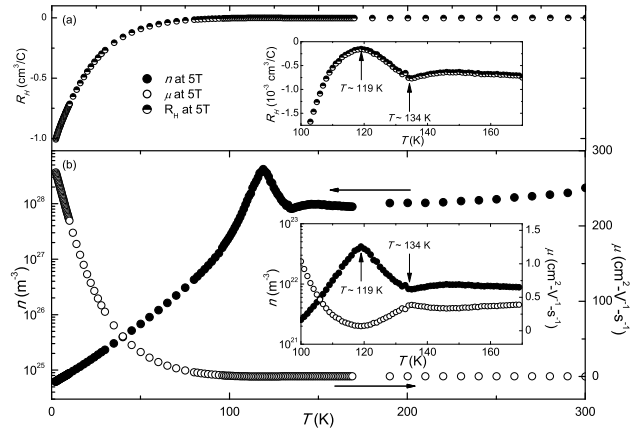}
    \caption{(a) Hall coefficient $R_{H}$ vs. temperature $T$ of single crystalline LaFeAsO in a magnetic field of 5 T applied perpendicular to the basal plane. The negative sign of $R_{H}$ indicates that the dominant charge carriers are electrons. (b) Charge carrier mobility $\mu$ and concentration $n$ for single crystalline LaFeAsO under 5 T of applied magnetic field applied perpendicular to the basal plane. The maximum in charge carrier concentration corresponds well with the feature in charge carrier mobility at the start of its low temperature upturn.}
    \label{fig:CAMLaFeAsORH02}
\end{center}
\end{figure}

Hall coefficient $R_{H}$ vs. temperature $T$ data for single crystalline LaFeAsO, measured using the van der Pauw configuration, are shown in Fig.~ \ref{fig:CAMLaFeAsORH02}. Similar to previous reports on polycrystalline samples of LaFeAsO,\cite{MCGUIRE01,KAMIHARA02} the sign of $R_{H}$ for the single crystal sample of LaFeAsO is negative from 2.2 K to room temperature, indicating that electrons are the predominant charge carriers.  Features in $R_{H}(T)$ at 119 K and 134 K correspond closely in temperature with features in $\rho(T)$ and $C(T)$. These features are likely associated with the SDW order and structural transition, respectively, as discussed earlier.

Mobility and concentration of charge carriers in single crystals of LaFeAsO were inferred from the Hall coefficient and magnetoresistance measurements through the relations $\mu = |R_{H}|/\rho(H)$ and $n = -1/eR_{H}$, respectively. These quantities, as defined, must be considered very carefully, because they assume a simple band structure which is probably inconsistent with the well established complexity of the band structure of this compensated material.\cite{KORSHUNOV02} Nonetheless, these quantities identify features associated with phase transitions of interest, as well as highlighting the aggregate behavior of the collection of charge carriers involved as shown in Fig.~\ref{fig:CAMLaFeAsORH02}. The charge carrier concentration and mobility are again consistent with the semiconductor-like behavior seen in the electrical resistivity and low density of electronic states inferred from the small value of $\gamma$ extracted from low temperature specific heat data.

\section{CONCLUDING REMARKS}

We have characterized the physical properties of single crystals of LaFeAsO grown in a NaAs flux by means of electrical transport, magnetization, specific heat, Hall coefficient, and magnetoresistance measurements. Two features seen in the first derivative of the electrical resistivity with respect to temperature correspond closely with features seen in specific heat measurements, and appear to be associated with the structural transition from tetragonal to orthorhombic symmetry at 134 K and the onset of SDW order at 114 K. X-ray diffraction measurements confirm that the structural transition occurs near 140 K. The structural transition temperature is reduced relative to the structural transition temperature of $\sim$ 156 K observed in polycrystalline samples of LaFeAsO.\cite{MCGUIRE01} The semiconducting-like behavior of the electrical resistivity of LaFeAsO single crystals can be fit with an Arrhenius law for temperatures above the onset of SDW order yielding an activated gap of $\Delta\sim9.5$ meV. This value for $\Delta$ is comparable to the high temperature data reported in the supplemental material of Ref.~\onlinecite{YAN01}.

Possible sources of the differences in lattice parameter values and transition temperatures, as well as the temperature dependence of the electrical resistivity between the single crystals of LaFeAsO studied in this work and polycrystals of LaFeAsO may be due to inclusions of other phases or atomic substituents that dope the single crystals with charge carriers. It is also possible that these discrepancies may be caused by defects or strain in the single crystals.  Additional studies are needed to explore these issues.

Electrical transport measurements were also performed under an applied magnetic field utilizing the van der Pauw method. From these data, we obtained the Hall coefficient $R_{H}$ and magnetoresistance. The sign of $R_{H}$ shows that the predominant charge carriers of this compensated metal are electrons, consistent with reports on polycrystalline samples.\cite{MCGUIRE01} Our data do not appear to show evidence for Dirac cones in the Fermi surface as previously reported\cite{PALLECCHI01} (there is no apparent linear component to the magnetoresistance curves). While the magnetoresistance data for these samples do not appear to obey Kohler's rule, we have shown that the data do collapse onto a single curve by normalizing both magnetic field and magnetoresistivity data relative to the maximum in each isotherm magnetic for fields up to 9 T and temperatures up to 130 K. We have also shown that there is a simple relationship between temperature and the limits of the quadratic magnetic field dependence of magnetoresistivity due to the cyclotron contribution.

\begin{acknowledgements}

\noindent Sample synthesis was funded by the US AFOSR-MURI (Grant FA9550-09-1-0603). Physical properties measurements were supported by the US DOE (Grant DE-FG02-04-ER46105). Research at Oak Ridge (MAM and BCS) was supported by the US Department of Energy, Basic Energy Sciences, Materials Sciences and Engineering Division.

\end{acknowledgements}

\clearpage

\bibliography{LaFeAsO}

\clearpage

\end{document}